\documentclass[a4paper]{jpconf}
\usepackage{graphicx}
\usepackage{amsmath}
\pdfoutput=1
\begin{document}
\title{Neutron detection in the SNO+ water phase}

%\author{Y Liu, on behalf of the SNO+ collaboration}
\author{Y Liu$^1$$^{,\ast}$, S Andringa$^2$, D Auty$^3$, F Bar$\tilde{\textbf{a}}$o$^2$, R Bayes$^4$, E Caden$^5$, C Grant$^6$, J Grove$^4$, B Krar$^1$, A LaTorre$^7$, L Lebanowski$^8$, J Lidgard$^9$, J Maneira$^2$, P Mekarski$^3$, S Nae$^2$, T Pershing$^{10}$, I Semenec$^1$, K Singh$^3$, P Skensved$^1$, B Tam$^1$ and A Wright$^1$ for the SNO+ collaboration}

\address{$^1$ Queen's University, Department of Physics, Engineering Physics \& Astronomy, Kingston ON K7L 3N6, Canada}
\address{$^2$ Laborat\'{o}rio de Instrumenta{\c c}$\tilde{a}$o e F{\'{i}}sica Experimental de Part{\'{i}}culas, Av. Prof. Gama Pinto, 2, 1649-003, Lisbon, Portugal}
\address{$^3$ University of Alberta, Department of Physics, 4-181 CCIS, Edmonton, AB T6G 2E, Canada}
\address{$^4$ Laurentian University, 935 Ramsey Lake Road, Sudbury, ON P3E 2C6, Canada}
\address{$^5$ SNOLAB, Creighton Mine \#9, 1039 Regional Road 24, Sudbury, ON P3Y 1N2, Canada}
\address{$^6$ Boston University, Department of Physics, Boston, MA 02215, USA}
\address{$^7$ University of Chicago, Department of Physics, Chicago, IL 60637, USA}
\address{$^8$ University of Pennsylvania, Department of Physics \& Astronomy, Philadelphia, PA 19104-6396, USA}
\address{$^9$ University of Oxford, The Denys Wilkinson Building, Keble Road, Oxford, OX1 3RH, UK}
\address{$^{10}$ University of California, 1 Shields Avenue, Davis, CA 95616, USA}

\ead{$^{\ast}$ yan.liu@owl.phy.queensu.ca}

\begin{abstract}
SNO+ is a multipurpose neutrino experiment located approximately 2 km underground in SNOLAB, Sudbury, Canada. The detector started taking physics data in May 2017 and is currently completing its first phase, as a pure water Cherenkov detector. The low trigger threshold of the SNO+ detector allows for a substantial neutron detection efficiency, as observed with a deployed $^{241}$Am$^{9}$Be source. Using a statistical analysis of one hour AmBe calibration data, we report a neutron capture constant of (208.2 $\pm$ 2.1$_{\textit{stat.}}$)$\mu$s and a lower bound of the neutron detection efficiency of 46\% at the center of the detector.
\end{abstract}

\section{Introduction}

SNO+\cite{Andringa:2015tza} is an underground multipurpose neutrino experiment that reuses the infrastructure of the SNO experiment. Located at a depth of 5890 meter water equivalent, the detector consists of a 12 m diameter spherical acrylic vessel (AV), which is currently filled with 1000 tonnes ultra pure water. Outside the AV, $\sim$9400 PMTs are mounted on an 18 m diameter geodesic stainless steel structure (PSUP). During the water phase, the experiment explores several interesting aspects of particle physics, including possible antineutrino searches using neutron tagging.

Electron antineutrinos are measured in most experiments via the inverse beta decay, as it provides a delayed neutron signal which can be used to suppress random background. In SNO+, neutrons propagating in light water will thermalize and then get captured by a proton, giving out a 2.2 MeV $\gamma$. Due to the small amount of energy deposited in the detector, neutron capture signals are very hard to detect in traditional pure water Cherenkov detectors. One approach\cite{Watanabe:2008ru} was explored in a previous experiment, which involves a forced trigger system after high energy events. Thanks to the good optical transparency and the upgraded electronics in SNO+, we are able to lower the detector threshold to $\sim$1.5 MeV, and can therefore trigger on the 2.2 MeV $\gamma$s with a substantial efficiency. In this proceeding we present the first result of neutron detection in a light water Cherenkov detector with normal trigger settings.

\section{AmBe calibration}

The $^{241}$Am$^{9}$Be source is a common neutron source that mimics the electron antineutrino signal very well. The $\alpha$-particle emitted by $^{241}$Am (with a half-life of 432.2 years) can be absorbed by the $^{9}$Be target, which will then decay into $^{12}$C through neutron emission. About 60\% of the time $^{12}$C is produced in an excited state and the immediate de-excitation predominantly emits a 4.4 MeV $\gamma$. The neutron, on the other hand, will thermalize and then get captured by a proton in $\sim$200 $\mu$s. The AmBe source provides coincident signals with the \textit{prompt} event being the 4.4 MeV $\gamma$ and the \textit{delayed} event being the neutron capture.

We used an AmBe source with a nominal strength of 1683.33 kBq and 62 Hz neutron rate\cite{jloach2009:diss}. The source was doubly encapsulated with black Delrin in SNO, and since then it had been used in the same form by other experiments at SNOLAB. Because the cleanliness of the existing encapsulations was not easy to assess, an extra encapsulation layer was designed, fabricated, leak tested and thoroughly cleaned prior to the source deployment in SNO+. This third encapsulation layer further reduces the risks of detector contamination.

\begin{figure}[h]
\includegraphics[width=14pc]{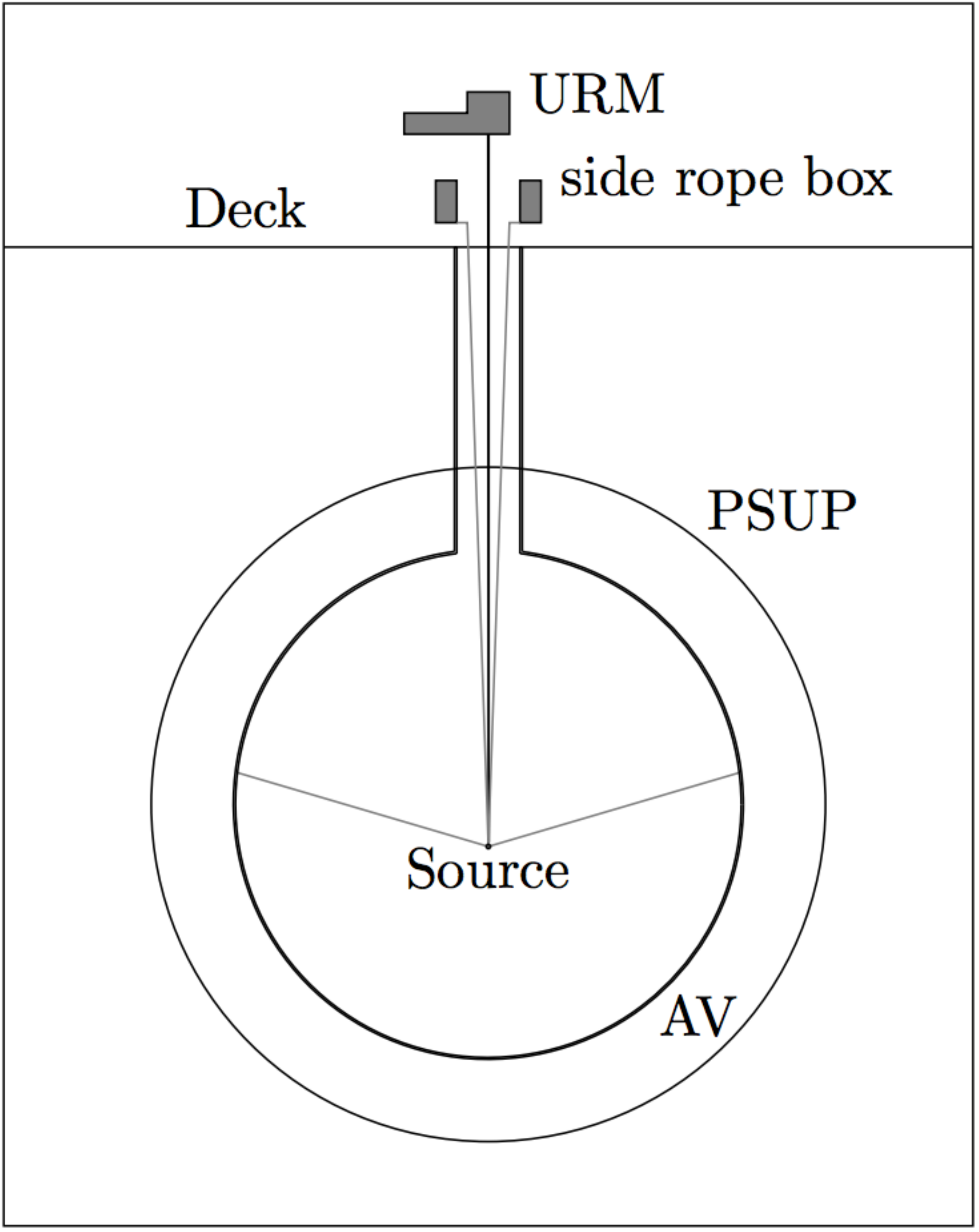}\hspace{2pc}%
\begin{minipage}[b]{18pc}\caption{\label{fig:1}A schematic of the Manipulator system in SNO+. Four upgraded side rope boxes are directly mounted on the UI. Calibration sources can be positioned in two vertical planes by adjusting the tensions and lengths of the ropes.}
\end{minipage}
\end{figure}

SNO+ reuses most of the SNO infrastructure for calibration\cite{Moffat:2005tq} in the water phase with major upgrades on the side rope boxes. The Manipulator system, show schematically in Figure \ref{fig:1}, contains an Umbilical Retrieval Mechanism (URM) which is mounted on the Universal Interface (UI). Calibration sources that are attached to a 33 meters umbilical and side ropes can then be placed in different positions inside the detector. 

The AmBe calibration was carried out in January 2018, at 23 different positions along the two axes: the vertical z-axis and the y-axis which points to the North. This allows detailed studies across the full detector volume. The total AmBe calibration time is 15 hours.

\section{AmBe data analysis}

Because the AmBe source is an untagged source, traditional analyses utilize time, position and other cuts to obtain a relatively pure sample. This method requires estimation of the background contamination, which often leads to large systematic uncertainties. The statistics can also be greatly reduced because of the stringent cuts. Here we present a statistical analysis that avoids the above problems, and also gives a direct measurement of the neutron detection efficiency.

The analysis started by filling a histogram with the time differences between the \textit{prompt} event and the \textit{delayed} event. These \textit{prompt} events and \textit{delayed} events were selected with a minimum Nhit (number of hit PMTs in one event) cut and only the first \textit{delayed} event was kept for each \textit{prompt} event. We did not exclude the possibility of one event being both the \textit{delayed} event in one pair and the \textit{prompt} event in the next pair. This histogram can then be categorized into three components:

\begin{itemize}
\item True-True event: the \textit{prompt} is the 4.4 MeV $\gamma$, and the \textit{delayed} is the associated neutron. 
\item True-Fake event: the \textit{prompt} is the 4.4 MeV $\gamma$, but the \textit{delayed} is background event.
\item Fake-Fake event: both \textit{prompt} and \textit{delayed} are backgrounds. The distribution of Fake-Fake events will follow an exponential with an exponential constant of the background rate.
\end{itemize}

For a True-Fake event, the prompt 4.4 MeV $\gamma$ is correctly identified. If the neutron does not trigger the detector or gets removed by the Nhit cut, the distribution will be an exponential with an exponential constant of the background rate. However, if the neutron triggers the detector and passes the Nhit cut, the \textit{delayed} event can be either the associated neutron or a background event that happens to appear before the neutron. Therefore, the probability of the \textit{delayed} event being a background is:
\begin{equation}
\begin{aligned}
\textrm{Prob}_{\gamma - b\textrm{ before n }}(\textrm{t}) & = P \cdot E \cdot R_{2} e^{-R_{2}\textrm{t}} \cdot (1- \int_0^{\textrm{t}} \lambda e^{-\lambda \textrm{t'}}\textrm{dt'}) \\
& = P \cdot E \cdot R_{2} e^{-(R_{2}+\lambda)\textrm{t}}
\end{aligned}
\end{equation}

where $P$ is the fraction of true 4.4 MeV $\gamma$s in the \textit{prompt} events, $E$ is the neutron detection efficiency, $\lambda$ is the neutron capture constant and $R_2$ is the background rate.

Similarly, for a True-True event we have:
\begin{equation}
\textrm{Prob}_{\gamma - n\textrm{ before b }}(\textrm{t}) = P \cdot E \cdot \lambda e^{-(\lambda+R_{2})\textrm{t}}
\end{equation}

\begin{figure}[h]
\begin{center}
\includegraphics[scale=0.5]{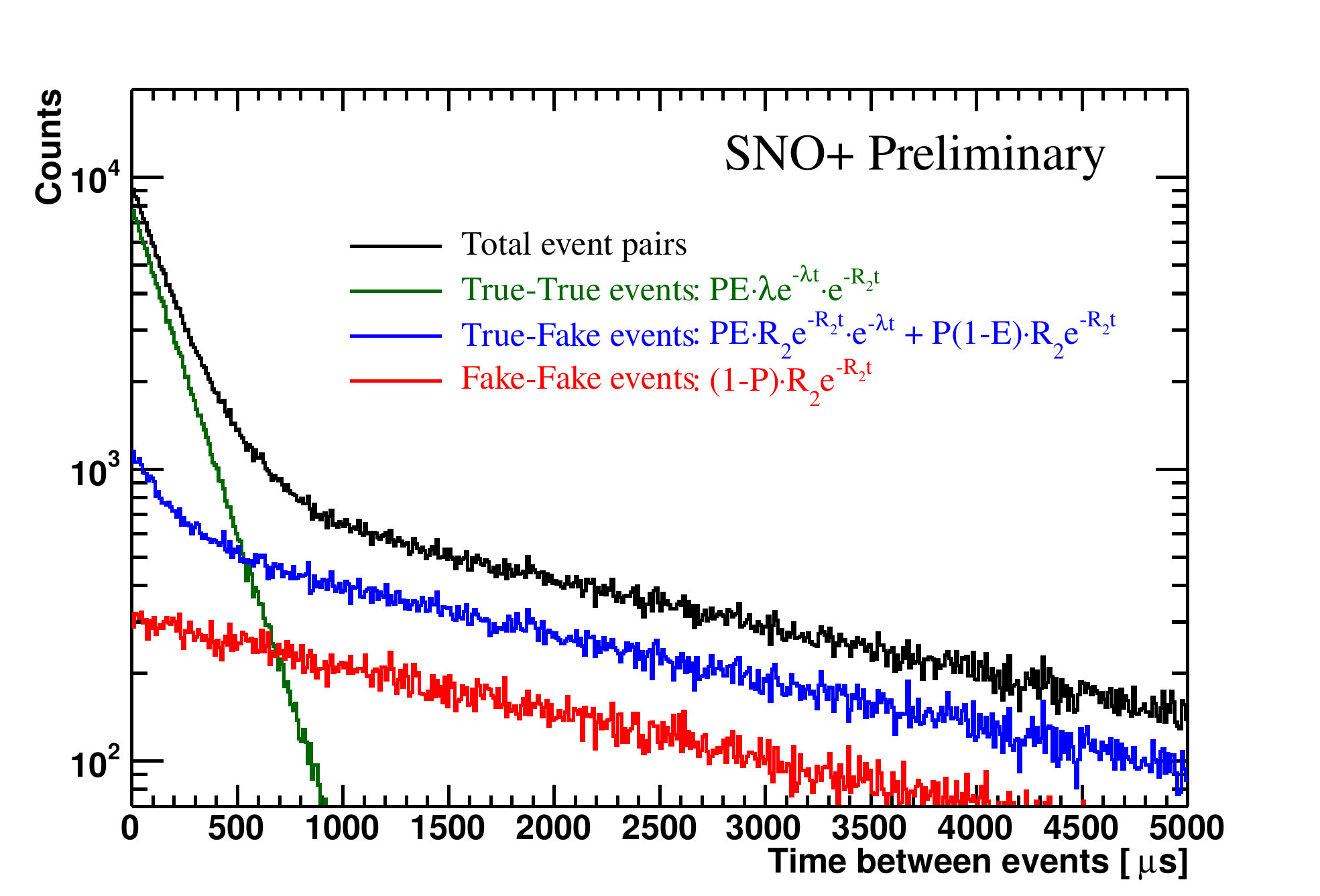}
\end{center}
\caption{\label{fig:2}Components of the time difference histogram from a toy MC model.}
\end{figure}

Therefore, we derived the fit function for the time difference histogram:

\begin{equation}
F(\textrm{t}) = N \cdot R_{1} (P \cdot E \cdot (\lambda+R_{2}) e^{-(\lambda+R_{2}) \textrm{t}} + (1-P \cdot E) \cdot R_{2} e^{-R_{2} \textrm{t}})
\end{equation}

where $N$ is a normalization factor associated with detector livetime and histogram bin size and $R_{1}$ is the \textit{prompt} event rate. Figure \ref{fig:2} shows the three different components of the time difference histogram from a toy MC model.

\section{Analysis results}

\begin{figure}[h]
\begin{center}
\includegraphics[scale=0.5]{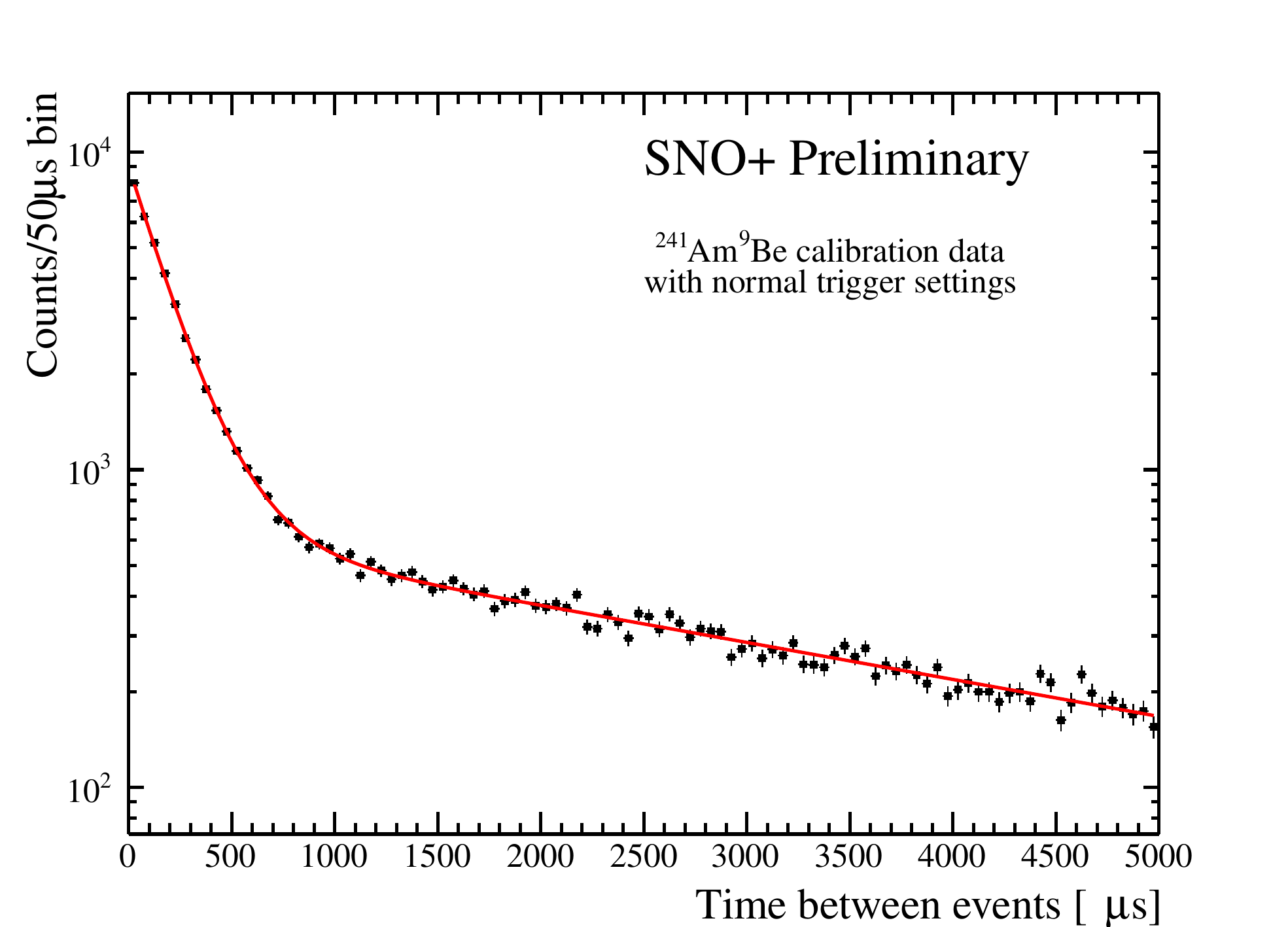}
\end{center}
\caption{\label{fig:3}Time difference histogram fitted with the analytical function. Shown here are data from a one hour central run.}
\end{figure}

We present here the analysis result for a one hour AmBe run where the source is positioned at the center of the detector. Figure \ref{fig:3} shows the fitted results. We report a neutron capture constant of (208.2 $\pm$ 2.1$_{\textit{stat.}}$)$\mu$s, which is consistent with previous measurements\cite{Super-Kamiokande:2015xra}\cite{Zhang:2013tua}\cite{Cokinos:1977zz}.

By varying the Nhit cut on the \textit{prompt} events, we found the maximum $P \cdot E$ value to be 46\%. By definition $P$ is smaller than 1,therefore we derived a conservative lower limit for $E$ at the center of the detector:

\begin{equation}
E > (P \cdot E)_{max} = 46\%
\end{equation}

This is the highest neutron detection efficiency achieved to date in a pure water Cherenkov detector.

\begin{figure}[h]
\begin{center}
\includegraphics[scale=0.5]{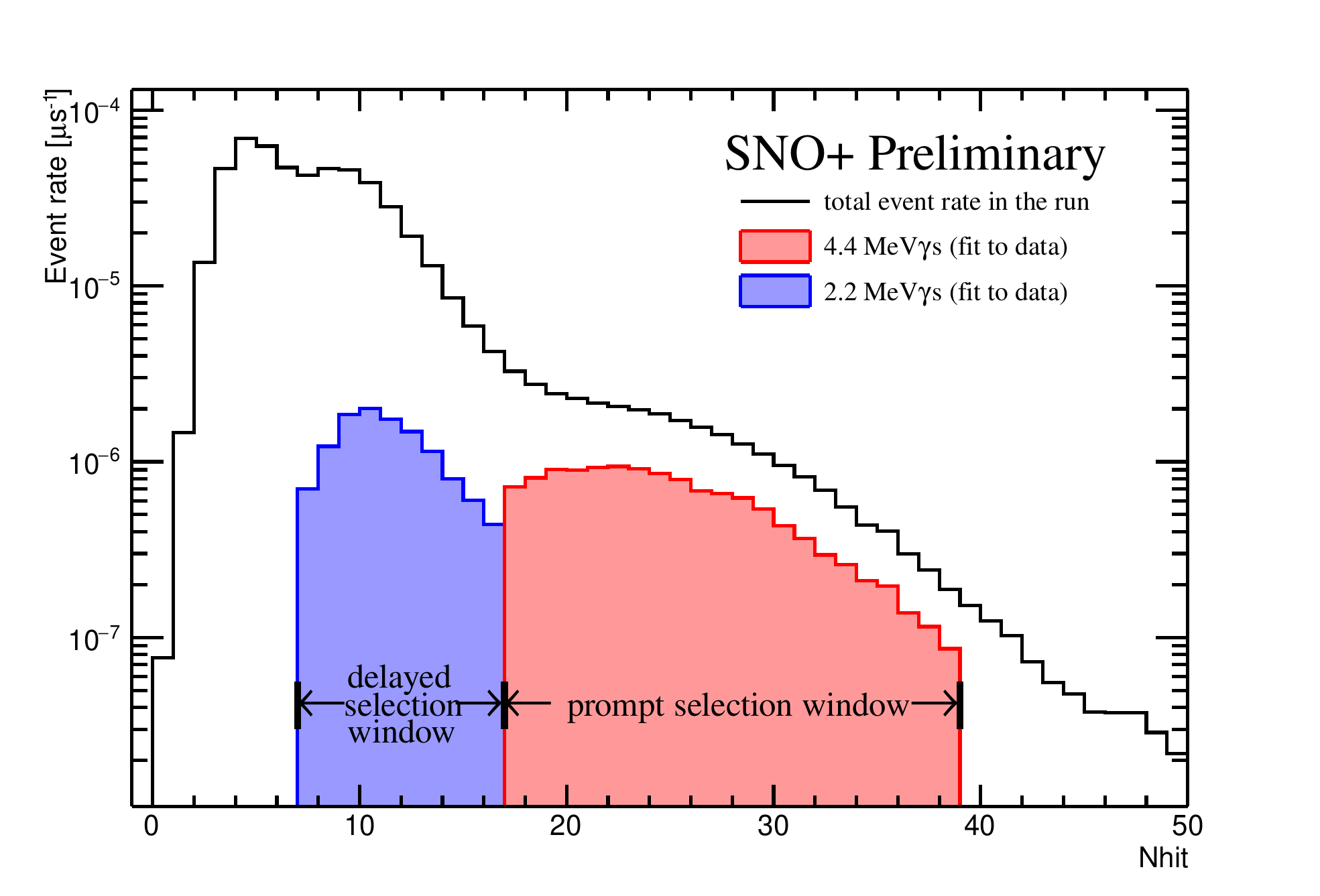}
\end{center}
\caption{\label{fig:4}Derived Nhit distributions for both the 4.4 MeV $\gamma$s and the 2.2 MeV $\gamma$s from neutron capture. The Nhit distributions are compared with the total event rate in this AmBe calibration run.}
\end{figure}

Note that $R_1 \cdot P \cdot E$ is the rate of true $\gamma$-n coincidences, regardless of whether there is a background event before the neutron. By calculating the difference of $R_1 \cdot P \cdot E$ for two consecutive Nhit cuts, we can plot the Nhit distributions for both the 4.4 MeV $\gamma$s and the neutrons, which is shown in Figure \ref{fig:4}. Also note that using this method, the Nhit distributions are not biased by random background and therefore can be used for energy calibration for the detector. It will not only provide an additional check on the energy scale linearity, but additionally will help to constrain measurement of backgrounds from U/Th chains. 

\section{Conclusion}

SNO+ started its water phase in May 2017 and has been steadily taking data since then. An AmBe source was deployed in the SNO+ detector for energy calibration and potential antineutrino searches. In this proceeding, we presented a novel data analysis method using statistical separation. With one hour central run data, we measured the neutron capture constant to be (208.2 $\pm$ 2.1$_{\textit{stat.}}$)$\mu$s. A lower limit of 46\% neutron detection efficiency was obtained at the SNO+ detector center. This is the highest neutron detection efficiency achieved to date in a pure water Cherenkov detector.

\ack{}

This work is supported by ASRIP, CIFAR, CFI, DF, DOE, ERC, FCT, FedNor, NSERC, NSF, Ontario MRI, Queen's University, STFC, UC Berkeley and benefitted from services provided by EGI, GridPP and Compute Canada. The poster presenter thanks FCT (Funda\c{c}$\tilde{a}$o para a Ci$\hat{e}$ncia e a Tecnologia, Portugal) and the Arthur B. McDonald Canadian Astroparticle Physics Research Institute for financial support. We thank SNOLAB and Vale for valuable support.

\section*{References}
\bibliographystyle{iopart-num}
\bibliography{ref}

\end{document}